\documentclass{iopart}

\usepackage[]{graphicx}
\usepackage{epsfig}

\begin{document}

\title{Photon polarisation entanglement from distant dipole sources}
\author{Yuan Liang Lim and Almut Beige}

\address{Blackett Laboratory, Imperial College London, Prince Consort Road, London SW7 2BZ, UK}

\begin{abstract}
It is commonly believed that photon polarisation entanglement can only be obtained via pair creation within the {\em same} source or via postselective measurements on photons that overlapped within their coherence time inside a linear optics setup. In contrast to this, we show here that polarisation entanglement can also be produced by {\em distant} single photon sources in free space and without the photons ever having to meet, if the detection of a photon does not reveal its origin -- the which way information. In the case of two sources, the entanglement arises under the condition of two emissions in certain spatial directions and leaves the dipoles in a maximally entangled state.
\end{abstract}

Secure quantum cryptographic protocols \cite{BB84} rely on the creation of entangled photon pairs or at least the presence of effective entanglement in the scheme \cite{entanglement}. In order to establish a shared secret key, the sender (Alice) produces a stream of photons that she sends to the receiver (Bob). Each photon should be prepared in a state known to Alice and encodes one random bit of information. The secret key is extracted from the outcomes of the measurements that Bob performs on the incoming photons. To prevent an eavesdropper from obtaining information about the key without being noticed, it is important that each bit is encoded in the state of only one photon \cite{single}. Other applications for single photon states can be found in linear optics quantum computing \cite{KLM}. 

Due to the variety of interesting applications, a lot of effort has been made in the last years to develop new and reliable photon sources. Each of them has its respective merits. Current single photon sources \cite{Gisin} include atom-cavity schemes \cite{Rempe} as well as quantum dots \cite{Yamamoto}, NV color centres in a diamond \cite{Weinfurter,Grangier} and tunable photonic band gap structures \cite{stefan}. It is commonly believed that  entangled photon pairs can only be created within the {\em same} source as in atomic cascades \cite{Aspect}, in parametric down conversion schemes \cite{parametric} and in the biexciton emission of a single quantum dot in a cavity \cite{milburn}. If the entanglement is not created within the same source, single photons can be brought together to overlap within their coherence time on a beamsplitter where a postselective entangling measurement has to be performed on the output ports \cite{yurke}.  

In contrast to this, we show that polarisation entanglement can also
be obtained when the photons are created by {\em distant}  sources
without the photons ever having to meet and without having to control
their emission times precisely. As an example we analyse the photon
emission from two dipole sources that might be realised in the form of
trapped atoms, diamond NV color centres, quantum dots or by using
single atoms doped onto a surface. An interaction between the sources
is not required. Each source should possess a $\Lambda$-type
three-level configuration with the two degenerate ground states $|0
\rangle$ and $|1 \rangle$, the exited state $|2 \rangle$ and optical transitions corresponding to the two orthogonal polarisations ``$+$" and ``$-$". Polarisation entanglement arises under the condition of the emission of two photons in different but carefully chosen directions independent from the initial state of the sources.

To understand how the proposed scheme works, it is important to recall that a detector always observes an integer number of photons. At the same time, fluorescence from two distant dipole sources can produce an interference pattern on a far away screen, if the distance of the screen from the sources is much larger than the distance between the sources \cite{Scully,Eichmann,Schon}. This wave-particle dualism implies that both sources contribute {\em coherently} to the creation of each photon. Consequently, the emission of one photon leaves a trace in the states of {\em all} its potential sources, depending on its polarisation and the direction of its wave vector \cite{Schon}, and can thus affect the state of the subsequently emitted photon.  

In the following, the detectors of Alice and Bob are placed such that all wave vector amplitudes contributing to the creation of a second photon with the same polarisation as the first one interfere destructively. In case of the collection of two photons (one by Alice and one by Bob) the shared pair has to be in a superposition of the state where Alice receives a photon with polarisation ``$+$" and  Bob a photon with polarisation ``$-$" and the state where Alice receives a photon  with polarisation ``$-$" and  Bob a photon with  polarisation ``$+$". Both share a maximally entangled pair, if the amplitudes for these two states are of the same size. In summary, polarisation entanglement is obtained with the help of postselection and interference effects. Related mechanisms have been proposed in the past to create atom-atom entanglement \cite{cabrillo}. 

The pair creation scheme proposed in this paper is feasible with present technology and might offer several advantages to quantum cryptography. In contrast to parametric down conversion \cite{Gisin}, the setup guarantees antibunching between subsequent photon pairs since the creation of a new pair is not possible without reexcitation of both sources. Furthermore, the scheme is robust. The final photon state does not depend on the initial state of the sources  in case of a successful collection. Another important advantage to note is that the scheme offers the possibility to generate {\em multiphoton} entanglement by incorporating more than two radiators in the setup \cite{next}.

Let us now discuss the creation of an entangled photon pair in detail. We describe the interaction of the dipole sources with the surrounding free radiation field by the Schr\"odinger equation.  The annihilation operator for a photon with wave vector ${\bf k}$, polarisation $\lambda$ and polarisation vector\footnote{In this paper, the notation is chosen such that $\hat {\bf x} \equiv {\bf x}/\| {\bf x} \| $.} $\hat {\bf \epsilon}_{\hat {\bf k}\lambda}$ is $a_{{\bf k}\lambda}$. The two dipole sources considered here are placed at the fixed positions ${\bf r}_1$ and ${\bf r}_2$ and should be identical in the sense that they have the same dipole moment ${\bf D}_{2j}$ for the 2-$j$ transition ($j=0,1$). The energy separation between the degenerate ground states and level 2 is $\hbar \omega_0$ while $\omega_k=kc$ and $L^3$ is the quantisation volume of the free radiation field. Using this notation, the system Hamiltonian becomes within the rotating wave approximation and with respect to the interaction-free Hamiltonian
\begin{equation} \label{21}
H_{\rm I} = \sum_{i=1,2} \sum_{j=0,1} \sum_{{\bf k},\lambda}  \hbar g_{{\bf k}\lambda}^{(j)} \, 
{\rm e}^{ -{\rm i} (\omega_0 - \omega_k) t} \,  {\rm e}^{ -{\rm i} {\bf k} \cdot {\bf r}_i} \, 
a_{{\bf k}\lambda}^\dagger \, |j \rangle_{ii} \langle 2| + {\rm H.c.} ~,
\end{equation}
where
\begin{eqnarray}
g_{{\bf k}\lambda}^{(j)} 
&=& {\rm i} e \, \left[ \frac{\omega_k}{2 \epsilon_0 \hbar L^3 } \right]^{1/2}
\! ({\bf D}_{2j} , \hat {\bf \epsilon}_{\hat {\bf k}\lambda}) 
\end{eqnarray}
is the coupling constant for the field mode $({\bf k},\lambda)$ to the 2-$j$ transition of each source. The rotating wave approximation corresponds to neglecting the non-energy conserving terms that describe the excitation of atoms combined with the creation of a photon or the deexcitation of atoms combined with the annihilation of a photon. These effects are not unphysical \cite{knight,he} but their contribution to the time evolution of the described system can be shown to be very small and almost impossible to observe.

To describe the effect of an emission on the state of the sources, we introduce the spontaneous decay rate of the 2-$j$ transition $\Gamma_j \equiv (e^2 \omega_0^3 \, |{\bf D}_{2j}|^2)/(3\pi \epsilon_0 \hbar c^3)$ and the reset operator $R_{{\rm X},\lambda}$. If $|\varphi \rangle$ is the state of the sources prior to an emission, it becomes $R_{{\rm X},\lambda} |\varphi \rangle/\| \cdot \|$ immediately afterwards if the created photon has polarisation $\lambda$ and a wave vector pointing in the $\hat {\bf k}_{\rm X}$ direction. Proceeding as in \cite{Schon}, $R_{{\rm X},\lambda}$ can be derived from the Hamiltonian (\ref{21}) and the projection postulate under the assumption of the detection of a one-photon state and is given by
\begin{eqnarray} \label{R21}
R_{{\rm X},\lambda} & \equiv & \sum_{i,j}  \left[ {3 \Gamma_j \over 8 \pi} \right]^{1/2}
(\hat{\bf D}_{2j} , \hat {\bf \epsilon}_{\hat{\bf k}_{\rm X} \lambda})  \, 
{\rm e}^{-{\rm i} k_0 \hat {\bf k}_{\rm X} \cdot {\bf r}_i} \,  |j \rangle_{ii} \langle 2| ~.~~
\end{eqnarray}
For convenience the reset operator has been defined such that $\| R_{{\rm X},\lambda} |\varphi \rangle \|^2$ gives the probability density for the emission of a photon with $\hat{\bf k}_{\rm X}$ and $\lambda$. 

The no-photon time evolution of the system, i.e.~the evolution between subsequent emissions, can be described by the quantum jump approach \cite{HeWi11}. This approach provides a non-Hermitian conditional Hamiltonian $H_{\rm cond}$ which also derives from the Hamiltonian (\ref{21}) under the assumption of environment-induced photon measurements.  For the setup considered here, one finds
\begin{eqnarray} \label{hcond}
H_{\rm cond} &=& -{\textstyle {{\rm i} \over 2}} 
\sum_{i=1,2}  \hbar (\Gamma_0 + \Gamma_1) \, |2 \rangle_{ii} \langle 2|  ~.
\end{eqnarray}
Given the initial state $|\varphi \rangle$, the state of the sources equals $U_{\rm cond}(t,0) |\varphi \rangle/\| \cdot \|$ at time $t$ under the condition of no detection. The non-unitary conditional time evolution operator  $U_{\rm cond}(t,0)$ has been derived such that $\| U_{\rm cond}(t,0) |\varphi \rangle \|^2$ is the corresponding no-photon probability. 

Let us now calculate the state of the system under the condition of the emission of two photons, the first one at $t_1$ in the $\hat{\bf k}_{\rm X}$ direction and the second one at $t_2$ in the  $\hat{\bf k}_{\rm Y}$ direction. If the initial state of the dipole sources at $t=0$ is $|\varphi_0 \rangle$, whilst the free radiation field is in its vaccum state and $|1_{{\rm X},\lambda} \rangle$, and $|1_{{\rm Y},\lambda'} \rangle$ denote normalised one-photon states with the parameters $\hat {\bf k}_{\rm X},\lambda$ and $\hat {\bf k}_{\rm Y},\lambda'$, the unnormalised state of the system equals  after the second emission
\begin{eqnarray} \label{evian}
|\psi (\hat{\bf k}_{\rm Y}, t_2 | \hat{\bf k}_{\rm X}, t_1) \rangle 
&=& \sum_{\lambda, \lambda'}  \,  |1_{{\rm Y},\lambda'} \rangle  |1_{{\rm X},\lambda} \rangle \otimes R_{{\rm Y},\lambda'} \, U_{\rm cond}(t_2-t_1,0) \, R_{{\rm X},\lambda} \nonumber \\
&& \times U_{\rm cond}(t_1,0) |\varphi_0 \rangle ~.~~
\end{eqnarray}
Note that $\| \, |\psi (\hat{\bf k}_{\rm Y}, t_2 | \hat{\bf k}_{\rm X}, t_1) \rangle \, \|^2$ yields the probability density for the corresponding event. 
        
To assure that Alice and Bob can receive a polarisation entangled pair, they should place their detectors in directions $\hat {\bf k}_{\rm A}$ and $\hat {\bf k}_{\rm B}$ with
\begin{equation} \label{hold}
{\rm e}^{ -{\rm i} k_0 \hat {\bf k}_{\rm A} \cdot  {\bf r}_1 } = {\rm e}^{ -{\rm i} k_0 \hat {\bf k}_{\rm A} \cdot {\bf r}_2 } 
~~ {\rm and} ~~ 
{\rm e}^{-{\rm i} k_0 \hat {\bf k}_{\rm B} \cdot  {\bf r}_1 } = - {\rm e}^{ -{\rm i} k_0 \hat {\bf k}_{\rm B} \cdot  {\bf r}_2 }~.
\end{equation}
These positions can easily be determined experimentally by continuously exciting the atoms with two laser fields and observing the far field interference pattern of the spontaneously emitted photons as in the two-atom double slit experiment by Eichmann et al.~\cite{Eichmann}. If a polarisation filter is used and photons of only one polarisation are detected, it can be shown that the position of Alice's detector corresponds to an intensity maxima and the position of Bob's detector to an intensity minima of the interference pattern due to a spatial antibunching effect \cite{Mandel}. Moreover, the spatial angles, in which conditions (\ref{hold}) are fulfilled to a very good approximation, have a larger tolerance, the smaller the distance between the atoms which can result in a relatively high photon pair collection rate. For a more detailed analysis of the underlying two-atom double slit experiment see Ref.~\cite{Schon}.

Inserting condition (\ref{hold}) for Alice's and Bob's detector position into Equation (\ref{evian}), one finds that the collection of two photons, one by Alice and one by Bob, prepares the system in the (unnormalised) state
\begin{eqnarray} \label{hurray}
&& |\psi (\hat{\bf k}_{\rm B}, t_2 | \hat{\bf k}_{\rm A}, t_1) \rangle = |\psi (\hat{\bf k}_{\rm A}, t_2 | \hat{\bf k}_{\rm B}, t_1) \rangle \nonumber \\
&& \hspace*{1cm} = \, {3\over 8 \pi} \, (2 \Gamma_0 \Gamma_1)^{1/2} \, {\rm e}^{ -{\textstyle {1\over 2}} (\Gamma_0+\Gamma_1)(t_1+t_2) } \, {\rm e}^{ -{\rm i} k_0 (\hat {\bf k}_{\rm A}+ \hat {\bf k}_{\rm B}) \cdot  {\bf r}_1 } \, \langle 22| \varphi_0 \rangle \nonumber \\
&&  \hspace*{1.45cm}  \sum_{\lambda, \lambda'} \left[ \big( \hat{\bf D}_{20} , \hat {\bf \epsilon}_{\hat {\bf k}_{\rm B} \lambda'} \big) \big( \hat{\bf D}_{21} , \hat {\bf \epsilon}_{\hat {\bf k}_{\rm A} \lambda} \big) - \big( \hat{\bf D}_{21} , \hat {\bf \epsilon}_{\hat{\bf k}_{\rm B} \lambda'} \big) \big( \hat{\bf D}_{20} , \hat {\bf \epsilon}_{\hat{\bf k}_{\rm A} \lambda} \big) \right] \nonumber \\
&&  \hspace*{1.45cm} \times |1_{{\rm B}, \lambda'} \rangle  |1_{{\rm A}, \lambda} \rangle \otimes  |a_{01} \rangle 
\end{eqnarray}
with $|a_{01} \rangle \equiv (|01 \rangle - |10 \rangle)/\sqrt{2}$. After two emissions, the dipole radiators are left in a maximally entangled state which is completely disentangled from the free radiation field.

\begin{figure}
\begin{center}
\epsfxsize7.0cm
\epsfbox{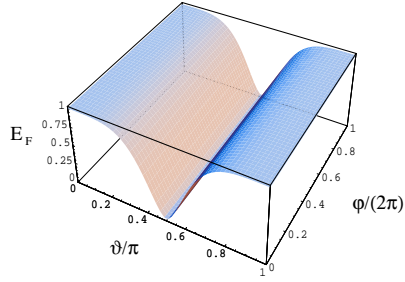} 
\end{center} 
\caption{The entanglement of formation $E_{\rm F}$ of the state
  $|\varphi_{\rm ph} \rangle$ of the shared photon pair as a function
  of the spherical coordinate $\vartheta$ and $\varphi$ of Bob's
  detector location while Alice collects photons in the $\hat {\bf
    z}$-direction.} \label{3d}
\end{figure}   

To calculate the final and normalised state of the system $|\varphi_{\rm ph} \rangle |a_{01} \rangle \equiv |\psi (\hat{\bf k}_{\rm B}, t_2 | \hat{\bf k}_{\rm A}, t_1) \rangle/\| \cdot \|$ explicitely, a coordinate system is introduced whose $\hat {\bf x}$-axis points in the direction of the line connecting the two sources and whose $\hat {\bf z}$-axis coincides with the quantisation axis. In addition, we choose $\hat {\bf k}_{\rm A} = (0,0,1)^{\rm T}$, $ \hat {\bf \epsilon}_{\hat{\bf k}_{\rm A} +} =  \hat{\bf D}_{20}  = (1,{\rm i},0)^{\rm T}/\sqrt{2}$ and $\hat {\bf \epsilon}_{\hat{\bf k}_{\rm A} -} = \hat{\bf D}_{21} = \hat{\bf D}_{20}^*$. Using the spherical coordinates $(\vartheta,\varphi)$ for Bob's detector position, one can write $ \hat {\bf \epsilon}_{\hat{\bf k}_{\rm B} +} = {\rm e}^{{\rm i} \varphi} ( \cos \vartheta \cos \varphi - {\rm i} \sin \varphi,  \cos \vartheta \sin \varphi  + {\rm i} \cos \varphi , - \sin \vartheta )^{\rm T}/\sqrt{2}$ and $\hat {\bf \epsilon}_{\hat{\bf k}_{\rm B} -} =  \hat {\bf \epsilon}_{\hat{\bf k}_{\rm B} +}^*$ which gives
\begin{eqnarray} \label{hurray!!}
|\varphi_{\rm ph} \rangle  &=& {1 \over 2  (1+\cos^2 \vartheta)^{1/2}} \, 
\big[  ( 1 + \cos \vartheta) \big(  \, |1_{{\rm B},+} \rangle  |1_{{\rm A},-} \rangle -  |1_{{\rm B},-} \rangle  |1_{{\rm A},+} \rangle  \, \big) \nonumber \\
&& + ( 1 - \cos \vartheta) \big(  \, {\rm e}^{2\rm i\varphi} \, |1_{{\rm B},+} \rangle  |1_{{\rm A},+} \rangle - {\rm e}^{-2\rm i\varphi} \, |1_{{\rm B},-} \rangle  |1_{{\rm A},-} \rangle  \, \big) \big] ~.~~
\end{eqnarray}
The entanglement of formation \cite{entropy} of $|\varphi_{\rm ph} \rangle$ as a function of Bob's detector position is $E_{\rm F} = -p \log_2 p - (1-p) \log_2 (1-p)$ with $p = \cos^2 \vartheta / (1 + \cos^2 \vartheta)$ and is shown in Figure \ref{3d}. The photon entanglement is close to unity, if $\hat {\bf k}_{\rm B}$ points in a direction relatively close to the quantisation axis $(\vartheta \approx 0)$ since this yields $(\hat{\bf D}_{20} , \hat {\bf \epsilon}_{\hat{\bf k}_{\rm X}+}) = (\hat{\bf D}_{21} , \hat {\bf \epsilon}_{\hat{\bf k}_{\rm X}-}) = 1$,  $(\hat{\bf D}_{20} , \hat {\bf \epsilon}_{\hat{\bf k}_{\rm X} -}) = (\hat{\bf D}_{21} , \hat {\bf \epsilon}_{\hat {\bf k}_{\rm X} +}) = 0$ (${\rm X}={\rm A},{\rm B}$) and 
\begin{eqnarray} \label{x}
|\varphi_{\rm ph} \rangle =( |1_{{\rm B},+} \rangle  |1_{{\rm A},-} \rangle -  |1_{{\rm B},-} \rangle  |1_{{\rm A},+} \rangle ) /\sqrt{2}
\end{eqnarray}
to a very good approximation. Photons with polarisation ``$+$" now
originate only from the 2-0 transition and photons with polarisation
``$-$" only from the  2-1 transition, as assumed in the beginning of
the manuscript. 

\begin{figure}
\begin{center}
\epsfxsize8.0cm
\epsfbox{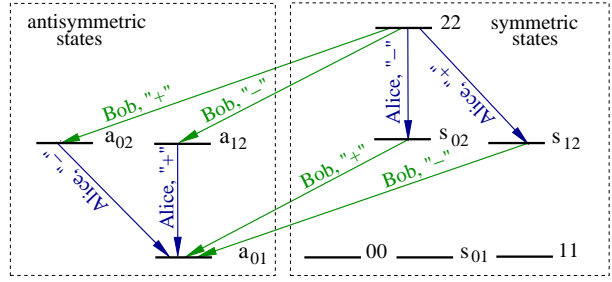} 
\end{center} 
\caption{Level configuration of the two atomic sources showing the relevant transitions that contribute to the creation of maximally entangled photon pair shared by Alice and Bob. Here $|s_{ij} \rangle$ and $|a_{ij} \rangle$ are the symmetric and antisymmetric Dicke states.} \label{level}
\end{figure}  

To give a more intuitive picture of the entanglement creation process, it is helpful to introduce the Dicke states $|s_{ij} \rangle \equiv ( |ij \rangle + |ji \rangle ) /\sqrt{2}$ and $|a_{ij} \rangle \equiv ( |ij\rangle - |ji\rangle)/\sqrt{2}$. Expressing the reset operators $R_{{\rm A},\lambda}$ and $R_{{\rm B},\lambda'}$ in this basis reveals that the detection of a photon by Alice transfers the source from a symmetric into a symmetric and an antisymmetric into an antisymmetric state while the detection of a photon by Bob results in a change of the symmetry of the state of the two sources. In addition, the condition of the emission of {\em two} photons implies that only the initial population in the state $|22 \rangle$ contributes to a successful outcome of the scheme. Taking this into account, Figure \ref{level} shows all the relevant transitions involved in the creation of the maximally entangled pair (\ref{x}). The reason for the final state of the sources not being entangled with the free radiation field is that there is only one unique antisymmetric ground state in the combined level scheme. 

An important fact to note is that the photon state (\ref{hurray!!}) does not depend on $t_1$ and $t_2$ and the order in which the photons arrive. This implies that Alice and Bob obtain an entangled pair irrespective of when the photons are collected. The efficiency of the scheme can therefore be calculated by integrating $\| \, |\psi (\hat{\bf k}_{\rm B}, t_2 | \hat{\bf k}_{\rm A}, t_1) \rangle \, \|^2 + \| \, |\psi (\hat{\bf k}_{\rm A}, t_2 | \hat{\bf k}_{\rm B}, t_1) \rangle \, \|^2$ over all times $t_1$ and $t_2$. This yields the probability $P$ for the collection of two photons within a time interval much larger than $1/(\Gamma_0 +\Gamma_1)$. Denoting the solid angle covered by Alice's and Bob's detector by $\Delta \Omega_{\rm A}$ and $\Delta \Omega_{\rm B}$ one finds
\begin{equation}
P = \left[{3 \over 8\pi} \right]^2 {2 \Gamma_0 \Gamma_1 \over (\Gamma_0+\Gamma_1)^2} \,  ( 1 + \cos^2 \vartheta) \,
| \langle 22 | \varphi_0 \rangle |^2  \, \Delta \Omega_{\rm A}  \Delta \Omega_{\rm B} 
\end{equation}
which is proportional to the initial population in $|22 \rangle$ and the solid angles $\Delta \Omega_{\rm A}$ and $\Delta \Omega_{\rm B}$. The smaller the distance between the sources, the easier it is to find relatively large areas where condition (\ref{hold}) holds to a very good approximation \cite{Schon}.

For applications like quantum cryptography it is useful to produce a stream of entangled photon pairs. The most convenient way to achieve this is to use continuous laser excitation. Before they start, Alice and Bob should agree about a time interval $\Delta T$ that assures that their measurement outcomes are highly correlated if both observe a click within $\Delta T$. The time $\Delta T$ has to be short compared to the time it takes to excite the atoms. On the other hand, $\Delta T$ should be much longer than $1/(\Gamma_0+\Gamma_1)$ to provide a reasonable efficiency of the proposed scheme. Antibunching guarantees that Alice and Bob can identify entangled pairs by comparing their photon arrival times at the end of each transmission via classical communication. 

As an example we describe now a setup for entangled photon pair creation with two trapped $^{87}$Rb atoms that is feasible with present technology \cite{rubidium}. The ground states $|0 \rangle$ and $|1 \rangle$ are obtained from the $5^2S_{1/2}$ levels with $F=1$ and have the quantum numbers $m_{\rm F}= -1$ and $m_{\rm F}= 1$. The excited state $|2 \rangle$ is provided by the $5^2P_{3/2}$ level with $F=0$. Suppose the atoms are initially in the $5^2S_{1/2}$ ground state with $F=1$ and $m_{\rm F}=0$ and a $\pi$ polarised laser field is applied to excite level 2. After spontaneous emission into the ground states $|0\rangle$ and $|1 \rangle$, another $\pi$ polarised laser reinitialises the system by coupling these states to the $5^2P_{3/2}$ states with $F=1$. From there the atoms return into the initial state via spontaneous decay. Due to their differences in polarisation and because of the detector locations, ``$+$" ($\sigma^+$) and ``$-$" ($\sigma^-$) polarised signal photons are distinguishable from laser photons and spontaneously emitted $\pi$ polarised photons. 

In conclusion, we proposed a scheme for the creation of polarisation entangled photon pairs by using two distant dipole radiators in free space. The entanglement is obtained by carefully choosing the detector positions with respect to the sources and arises under the condition of the collection of two photons independent of their emission times and the initial state of the sources. 
Another application of the scheme would be to prepare two distant dipole sources in the maximally entangled ground state $|a_{01} \rangle$. The presented idea might find interesting applications in quantum computing with trapped atoms, diamond NV color centres, quantum dots or single atoms doped onto a surface and opens new possibilities for the creation of antibunched polarisation entangled photon pairs and even multiphoton entanglement by including more than two radiators in the setup \cite{next}.

\ack
We are very grateful for valuable discussions with Dan Collins, Philippe Grangier and Axel Kuhn. Y. L. L. acknowledges funding from the DSO National Laboratories in Singapore. A. B. thanks the Royal Society and the GCHQ for funding as a James Ellis University Research Fellow. This work was supported in part by the European Union and the UK Engineering and Physical Sciences Research Council.

\end{document}